\documentclass[12pt]{spieman}  
\usepackage{color,soul}
\usepackage{url}
\usepackage{siunitx}
\usepackage{longtable}
\usepackage{comment}
\usepackage{amsmath,amsfonts,amssymb}
\usepackage{graphicx}
\usepackage{setspace}
\usepackage{tocloft}
\usepackage[left]{lineno}
\usepackage{caption}
\usepackage{subcaption}
\usepackage{lscape}
\usepackage[multiple]{footmisc}
\usepackage{array}
\title{Precision wavelength calibration for Radial Velocity measurements using Uranium lines between 3800-6900 \AA}
\author[a,*]{Rishikesh Sharma}
\author[a]{{Abhijit Chakraborty}}
\affil[a]{\rm Astronomy \& Astrophysics Division, Physical Research Laboratory, Ahmedabad 380009, India}

\cftpagenumbersoff{figure}
\cftpagenumbersoff{table}

\begin{document}
\maketitle
\graphicspath{{./}{figures/}}
\begin{abstract}
\noindent 
We present here the precise wavelength calibration of a high-resolution spectrum using Uranium (U) lines in the wavelength range of 3809 - 6833 \AA\ for precision radial velocity measurements for exoplanet detection or related astrophysical sciences. {We identify 1540 well-resolved U lines from a high-resolution (R=67,000) spectrum of the uranium-argon hollow cathode lamp (UAr HCL) using PARAS spectrograph in the aforesaid wavelength range.}
We calculate the neutral and first allowed transitions (Ritz wavelength) of U from its known energy levels and compare them with our observed central wavelengths. {We measure an offset of -0.15 m\AA\space in our final U line list.} The line list has an average measurement uncertainty of 15 m s$^{-1}$ (0.013 pixels or 0.28 m\AA). We included these lines to the PARAS data analysis framework to perform the wavelength calibration and then calculate the multi-order Radial Velocity (RV) of PARAS spectra. The typical dispersion of residuals around the wavelength solution of a UAr spectrum, using U lines, is found to be 0.8 m\AA\space($\sim$45 m s$^{-1}$). With the use of this line list, we {present} our results for the precision RV of an on-sky source (A RV standard star), and an off-sky source (A HCL) observed with PARAS along with UAr HCL. We measure the dispersion in absolute drift difference between two fibers (inter-fiber drift) for a span of 6.5 hours to be 88 cm s$^{-1}$, and the {RV dispersion} (${\sigma_{RV}}$) for a RV standard star, HD55575 over the course of $\sim$450 days to be 3.2 m s$^{-1}$. {These results are in good agreement with the previous ones measured using the ThAr HCL. It proves that the ThAr HCL with $\sim$ 99\% pure-Th are replaceable with the UAr HCL for the wavelength calibration of the high-resolution spectrographs such as PARAS (R $\leq$ 67,000) to achieve a RV precision of 1-3 m s$^{-1}$} in the visible region.\\ 
\end{abstract}


{\noindent \footnotesize{*}Rishikesh Sharma,  \linkable{rishikesh@prl.res.in} }

\section{Introduction}
\label{sec:intro}
The sudden increase in the population of exoplanet {candidates in the last decades from various space-based (like CoRoT\cite{Rouan1998}, Kepler \& K2\cite{Borucki2010}, TESS\cite{Ricker2015}, etc.) and ground-based transit surveys (like SuperWASP\cite{Swasp}, KELT\cite{KELT}, etc.) has opened a quest among the scientific community to build the high-precise radial velocity (RV) spectrographs, which can go down to sub - 1 m s$^{-1}$ of RV precision and eventually, detect and confirm the exoplanet candidates.} One of the main challenging aspects {for} such instruments is the wavelength calibration technique, which is evolved from the initial and established techniques of iodine cell and hollow cathode lamps (HCLs) to Fabry-Perot etalon and highly precise high-frequency laser combs. The laser frequency combs (LFCs) used in various spectrographs like {EXPRES \cite{EXPRES2016}} ESPRESSO \cite{espresso2014}, HARPS \cite{wilken2012}, HPF \cite{hpf2019}, etc., are the latest and most precise equipment for the wavelength calibration {that have} shown upto 2.5 cm s$^{-1}$ of RV precision on HARPS \cite{wilken2012} over a short-term (2 hours)\cite{Curto2012}. Besides being the most advance system for wavelength calibration, the use of LFCs is limited {due to their cost and complexity}. Alternatively, the conventional system for calibration, like HCLs\cite{Kerber2007} (e.g., ThAr), is {widely} used in the astronomical spectrographs due to their {relatively long life}, easy handling, simple structure, and less maintenance. Th metal being a single isotope, having narrow lines and a dense spectrum, is  used for the wavelength calibration of the astronomical spectrographs. PARAS (PRL Advanced Radial-velocity Abu-sky Search) \cite{paras2014,epic201} is a precision radial velocity fiber-fed, temperature and pressure-controlled spectrograph connected to a 1.2m telescope at PRL Mt. Abu Observatory in Gurushikar, Mt. Abu, India. {The high-precision RV spectrographs like HARPS and PARAS have used Th-Ar HCL for wavelength calibrations and shown long-term stability up to {0.8 m s$^{-1}$ \cite{HARPS_longterm}}  and 1 m s$^{-1}$\cite{paras2014}, respectively. The long-term stability here refers to the RV stability achieved over the course of {3} months or more.}\\
The ThAr HCLs from {S \& J Juniper \& Co}\footnote{\url{http://www.sjjuniper.com}} are currently used in many precision Radial-Velocity spectrographs. However, these HCLs with a cathode made of pure Th (up to 99\% purity) are no more commercially available; instead {they are available with a cathode made of Thorium-Oxide (ThO)}. \citenum{Nave2018} confirmed that this oxide impurity is responsible for the presence of molecular oxide bands and other unwanted features in the HCL's spectra. Thus, it becomes difficult to identify the faint Th atomic lines in the oxide bands and make them useless for wavelength calibration in some regions, making it tough to precisely characterize and track the Th lines. It degrades the precision of wavelength calibration of high-resolution spectra and, ultimately the RV precision. It becomes paramount to choose the proper element, which can replace the Th element in the HCLs and equally have a vast number of transitions to provide precise wavelength calibration for the astronomical spectrographs. \\
{We looked into the actinide series of the periodic table and found that U fits closest to the Th, and U HCLs are also readily available in the commercial market.} The U HCLs are available with cathode made of natural U. Unlike Th, the U atom is found in three isotopic forms in nature namely U$^{238}$, U$^{235}$, and U$^{234}$, with abundances of $\sim$99.275\%, $\sim$0.720\%, and $\sim$0.005\%, respectively. \citenum{Redman2012} showed that the second most abundant isotope has a hyper-fine structure, and thus its features spread over many lines with very less intensity and won't cause any significant problem in RV precision. Even, the recent study by \citenum{carmenes18} (Here on, S18) showed that the U could replace the Th in ThAr and Thorium-Neon (Th-Ne) HCLs for the wavelength calibration in the visible and NIR range more precisely as U has more number of lines than Th in both the wavelength regions. However, their line list starts from mid-visible and stretches up to NIR, i.e.,\ in the wavelength range of 5000-17000 \AA, which does not fully cover the PARAS wavelength range of 3809-6833 \AA. \citenum{palmer80} (here on, P80) was also previously published the U line list in the wavelength range of 3846-9091 \AA. We acknowledge both previous works of P80 and S18. The P80 line list covers the whole range, which is used for the wavelength calibration in PARAS spectra and hence, more suitable for comparison with our line list. Despite having more lines than Th in the visible range, U lines have never been checked a priori for long-term stability with high-resolution spectrographs. \\
{This work is primarily focuses on the precise wavelength calibration of an astronomical spectrograph like PARAS using the U lines in the visible band (3809 \AA \ - 6833 \AA) for radial velocity precision of 1-3 m s$^{-1}$ for exoplanet detection, characterization and, related astrophysical sciences. Here, we measure the radial velocities of a well-known RV standard star HD 55575 over a period of 450 days in order to check the feasibility of using U lines instead of Th lines for precision RV measurements of stars.} \\
We describe the observations with PARAS in section 2, while section 3 focuses on the data reduction and wavelength calibration of the UAr spectrum for identifying the lines. Section 4 elaborates the algorithms used for identifying the line features from a UAr spectrum and various steps involved in selecting the final line list, including those from U I and U II transitions. In section 5, we compare the observed central wavelength of selected lines and the calculated Ritz wavelength and discuss the final line list. In section 6, we incorporate the U line list in the PARAS wavelength calibration framework to do the RV measurements on a RV standard star HD55575 and show the on-sky performance for 450 days for the long-term stability of more than a year. Section 7 describes the possible future works, while section 8 summarizes our results.
{\section{PARAS Observations with UAr HCL}\label{sec:observations}
\subsection{Experimental set-up}}
We acquired the high-resolution spectra of the UAr HCL using the PARAS spectrograph\cite{paras2014} coupled with the 1.2m telescope at Gurushikhar Observatory, Mt Abu, India. The UAr HCL from {\tt PHOTRAN PTY LTD.} is used for the experiment. The spectrograph works at a resolving power of $\sim$ 67000 at the blaze peak wavelength of 5500 \AA. It is a white-pupil configuration fiber-fed spectrograph kept inside a thermally controlled vacuum chamber. {There are two fibers, namely star-fiber (or A fiber) and calibration-fiber (or B fiber) that feed the starlight and calibration lamp light into the spectrograph, respectively.} Inside the vacuum chamber, the light beam from both the fibers is collimated by an off-axis parabolic mirror, then dispersed by an R4 echelle grating, and then cross-dispersed by a prism, re-imaged onto the 4k X 4K e2v CCD, which has a pixel size of 15 $\mu$m. The details of the spectrograph are given in \citenum{paras2014}. The CCD detector has a sampling of $\sim$ 4 pixels and an average dispersion of 0.021 $\AA$ per pixel. It implies that each pixel corresponds to a velocity of 1145 \si{\meter\per\second}. {The previous study by \citenum{Kerber2007} has represented that a HCL's life and intensity of the emitted lines are strongly dependent on the operating current (OC). {The increase in OC increases the line intensities and number of transitions but reduces the HCL's life.} We set the OC to 10 mA to have a sufficient number of U transitions and, at the same time, {to increase the HCL's life}. We use the C610 HCL power supply from {\tt Cathodeon}\footnote{\url{https://www.msscientific.de/hollow_cathode_lamp_power_supply.htm}}, which provides the current stability up to 0.05\% after the warm-up time of 15 min, throughout the observations. We note here that the HCL was kept switched on throughout the night of observation.}
{\subsection{Off-sky Observations}}\label{sec:offsky}
Initially, we acquired a spectrum of the UAr HCL by illuminating its light in the star-fiber (or A-fiber) to identify the U emission lines. In order to do the initial wavelength calibration of the acquired UAr spectrum, a ThAr spectrum was also acquired in the same way just before the UAr spectrum. The spectrograph stability is such that the drift between the ThAr and the UAr spectrum is negligible for our purpose. 
{We did not use a neutral density filter for off-sky observations, and} set the exposure time at 400 sec to have the high SNR spectra {as well as to make sure that the high-intensity lines should not saturate the CCD}. \\
We also acquired a series of 400s UAr exposures {by simultaneously illuminating the UAr HCL's light in both the fibers (A and B fibers) for $\sim$ 6.5 hours}, to track each line's stability and calculate the {instrumental drifts} for {the mentioned time interval}.\\

{\subsection{On-sky Observations}}
{We observed a RV standard star HD55575, for the span of 450 days, acquiring 27 spectra during the period. These observations were done by illuminating the A-fiber with the starlight and simultaneously illuminating the B-fiber with the UAr light for precise wavelength calibration. The exposure time for these observations was set to 1200 sec, and the SNR ranges between 70-80 at 5500\AA. We use the neutral density filter (FSQ-OD50 from Newport) \footnote{\url{https://www.newport.com/p/FSQ-OD50##mz-expanded-view-462354496971}} with the UAr HCL for these long exposures to ensure that the high-intensity lines should not saturate the CCD. These spectra are used to calculate the RV precision achievable on stars with UAr HCLs (see Sec ~ \ref{sec:res_star}).}\\

\section{Data Reduction and Analysis}\label{sec:data_reduction}
We used the custom-designed automated pipeline\cite{paras2014} to extract the U spectra from PARAS 2-D (2 dimensional) images. The pipeline is based on the REDUCE package of the \citenum{reduce}, which is an optimal extraction code to extract the cross-dispersed echelle spectra. {We use the existing ThAr wavelength calibration process to do the initial wavelength calibration of the UAr spectrum, as described in \citenum{paras2014}. The ThAr spectrum, which was acquired just before the UAr spectrum as discussed in sec~\ref{sec:offsky}, is wavelength calibrated using this existing process in the PARAS pipeline. The pipeline uses a template of Th lines for the wavelength calibration. A template is defined here as an order-wise list of central wavelengths of Th lines to use for wavelength calibration. The template has a total of around 1000 Th lines with at least 8 lines per order.} {For a particular order} in {the ThAr} spectrum, the pixel position of {the central wavelength} for each line is found, and then a Gaussian is fitted to precisely {determine} the line position {in terms of pixels}. These fitted pixel positions and the line's central wavelengths for that order are then {fit with} a {third-order polynomial}. The resultant polynomial is used as the wavelength solution for that order. In this way, the wavelength calibration is done for all the 70 orders. This ThAr wavelength solution is applied over the UAr spectrum which was acquired for the line identification process.

\section{Identification and selection of U lines }\label{sec:id_lines}

\begin{figure}[!ht]
\centering
\includegraphics[width=0.7\textwidth]{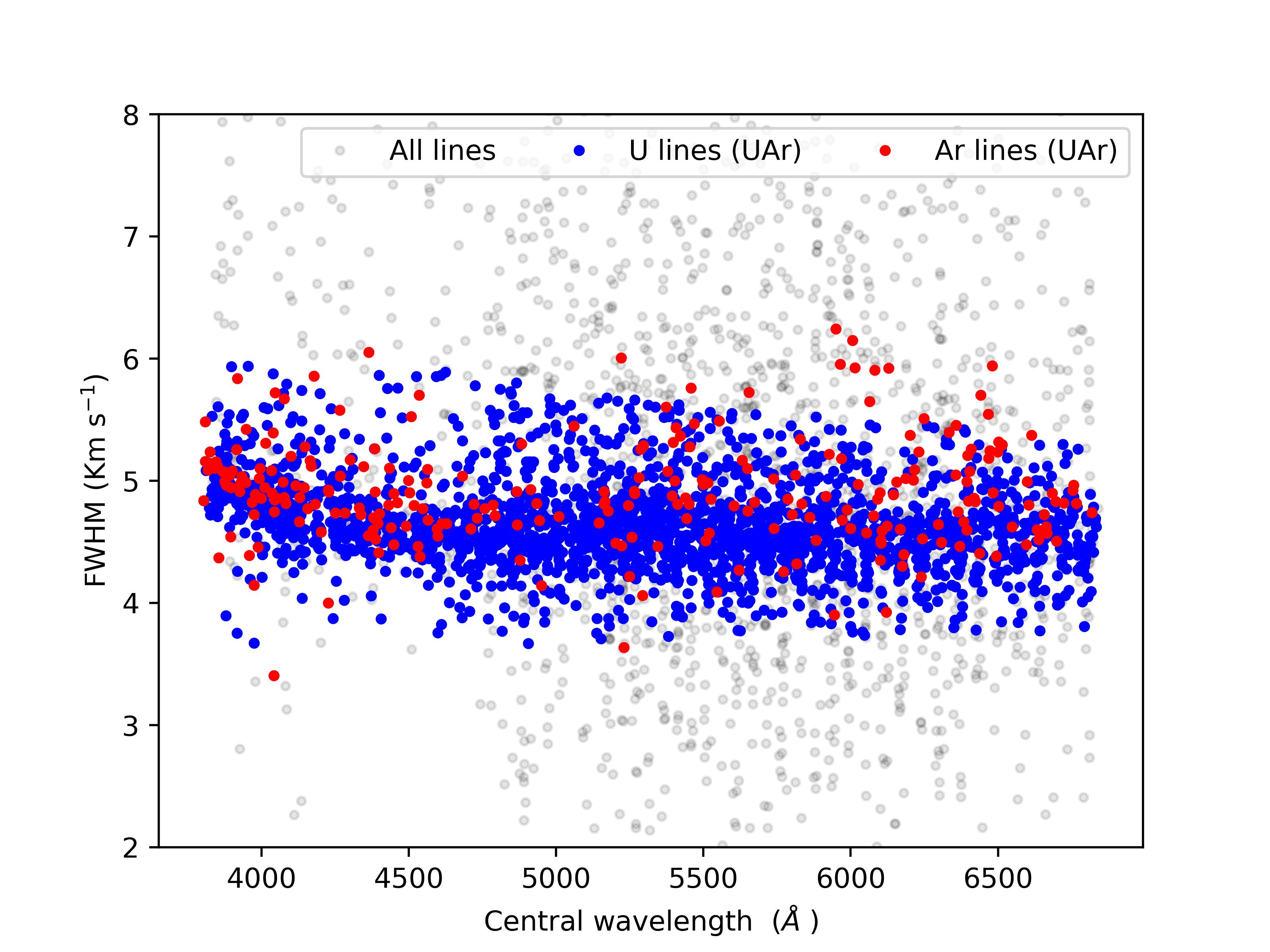}
\caption{{The FWHM of all the 4500 emission features identified in the first step (see Sec~\ref{sec:id_lines}) is plotted here in grey color. Overplotted in red are the widths of 270 Ar lines identified in Sec~\ref{sec:id_lines}. The blue dots represent the line widths of the 1819 lines of the preliminary U line list (see Sec~\ref{sec:id_lines}). It can be seen that the selected U lines and identified Ar lines are consistent with the instrumental resolution and have line widths between 3.75 - 6.0 Km s$^{-1}$.}}

\label{fig:fwhm_all}
\end{figure}
The wavelength calibrated UAr spectrum is now used for the line identification and selection process. The echelle spectra have overlapping wavelength regions over the two consecutive orders. We stitched the {UAr spectrum} distributed among various echelle orders for line identification and characterization by weighing the overlapping regions in the two consecutive orders according to their S/N ratio. This stitched spectrum is further used for the line identification process.\\
First, we identified all the {line features} present in the stitched UAr spectra, based on the slope of the spectra at {each point}, considering only those features with S/N ratio $\geq$ 30, to minimize the uncertainties in finding the central wavelength $(\sigma_{\lambda} < 1 m\AA)$. {We characterized each of these identified lines by fitting them with the {\tt MPFITPEAK} \cite{mpfit2009} Gaussian function, written in {\tt IDL}, and determined the line parameters. We have identified nearly 4500 emission features in this step. The intensities of these lines spread across the whole dynamic range of the CCD detector, and 3 high-intensity lines among them falling at 5915.3867, 6449.1625 \& 6826.9185 \AA, \space are found to be saturated. These lines are not considered for further selection and are excluded from the identified lines.}\\
\begin{figure}[!ht]
\centering
\includegraphics[width=0.7\textwidth, trim= {0.0cm 0.0cm 0.0cm 0.0cm}]{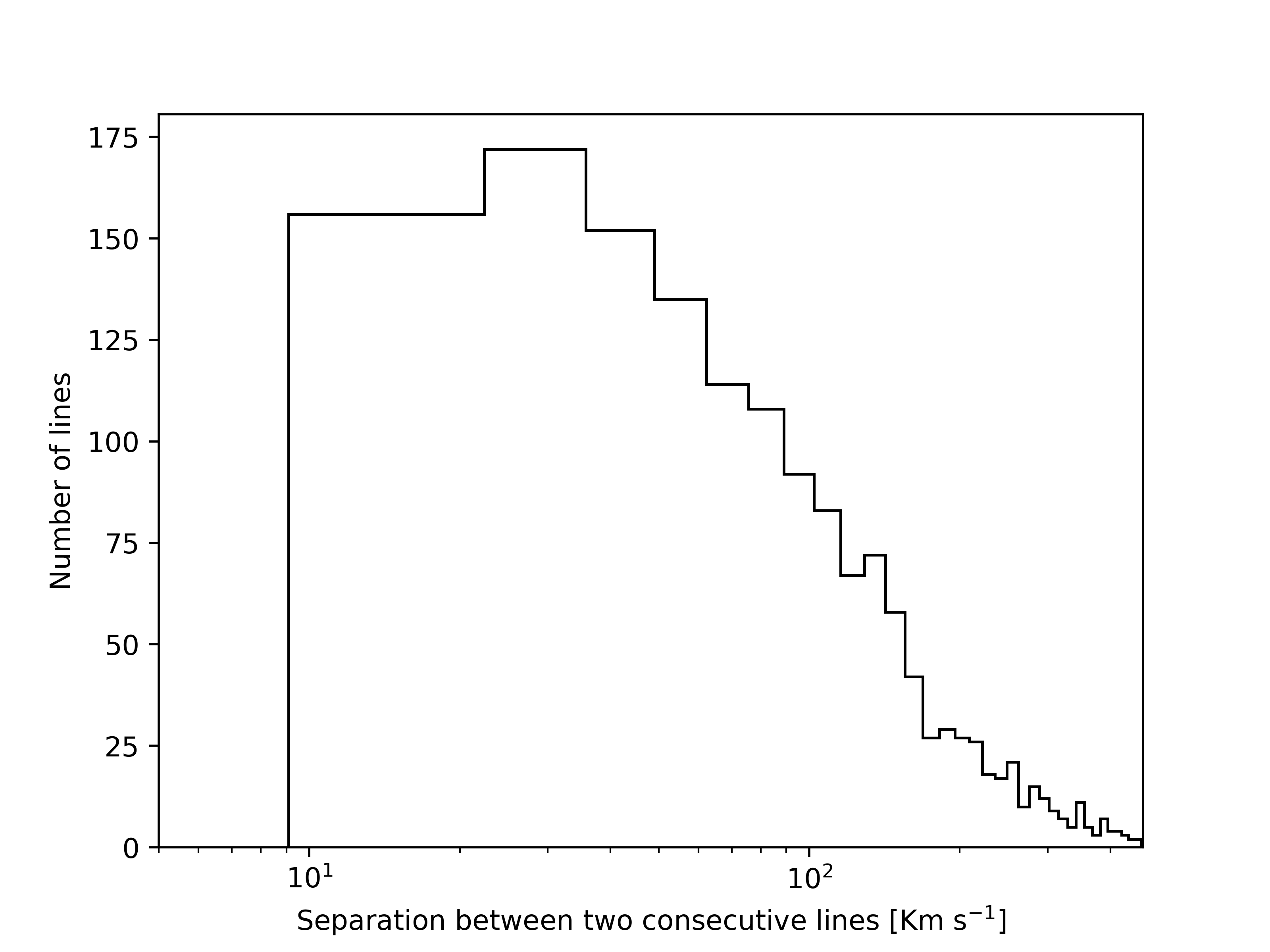}
\caption{The histogram plotted here shows the separation in Km s$^{-1}$ between two consecutive lines from our final line list. {The minimum separation is $\sim$ 9.0 Km s$^{-1}$, which is nearly the two resolution element PARAS, and shows that the lines in our line list are well-separated. The x-axis is in the log scale.}} \label{fig:hist_uar}
\end{figure}
The {lines} identified in the last step are {supposed to be emitted by either cathode filling metal U or the filling gas Ar. It is well-known that the Ar lines are not suitable for the precise wavelength calibration and can show wavelength drifts of tens of m s$^{-1}$ \cite{lovispepe2007}. The next step in the U line selection procedure is to remove the Ar lines.} So, we identified the Argon lines using the \citenum{lovispepe2007} ThAr line list, based on the fitted central wavelength of each line's Gaussian profile and eliminated them for the further selection process. \citenum{lovispepe2007} used the HARPS spectrograph for their work. Since HARPS has a higher resolution than PARAS, all the Argon lines from our spectra must be listed in the cited line list. We could identify 270 Ar lines in the UAr spectrum and removed them from the lines identified in the last step. { We have plotted the line widths (FWHM of the Gaussian) of Ar lines in red color in Fig~\ref{fig:fwhm_all}. It can be seen that their line widths are consistent with the PARAS instrumental resolution.}
The blended lines present in the UAr spectra can induce the centroid shifts much larger than the measurement errors associated with line positions. Therefore, It is better to remove the possible blends from the remaining lines. We used the line width of the remaining lines to identify the blends. Since the instrument works at a certain resolution (R $\sim$ 67,000; 4.5 Km s$^{-1}$), we only considered those lines whose widths are consistent with the instrumental resolution and lies within 3.75 - 6.0 Km s$^{-1}$. This eliminates the possible dangerous blends in the UAr spectra to be included in our further selection of the lines. In fig~\ref{fig:fwhm_all}, we have plotted the line widths of all the identified lines in grey color. The lines which pass the above selection criteria have been highlighted in blue. In fig~\ref{fig:fwhm_all}, few lines among 4500 lines with significantly smaller line widths ($\leq$ 3.75 Km s$^{-1}$) can be seen, which could be due to some artifacts or bad pixels and hot pixels in the CCD detector. As a precaution, we also removed those lines which appear single in our spectra but have other line features within, listed in the P80 line list, due to a much higher resolution of the instrument used by them (R = 600,000).

Since our main aim is to do the precise wavelength calibration of the spectrograph using UAr spectra, we strictly consider the lines that are well resolved or well-separated from their neighbors to fit a single-peaked 1-D (1-dimensional) Gaussian. This implies that the two neighboring lines with least separation of 2-resolution elements are considered for further selection. We have plotted a histogram (see fig-\ref{fig:hist_uar}), which shows the distance between two consecutive lines from the final line list. The minimum separation between two lines is 9.0 km s$^{-1}$, which nearly equal to 2-resolution elements of PARAS. 

\begin{figure}[!ht]
\centering
\includegraphics[width=0.7\textwidth]{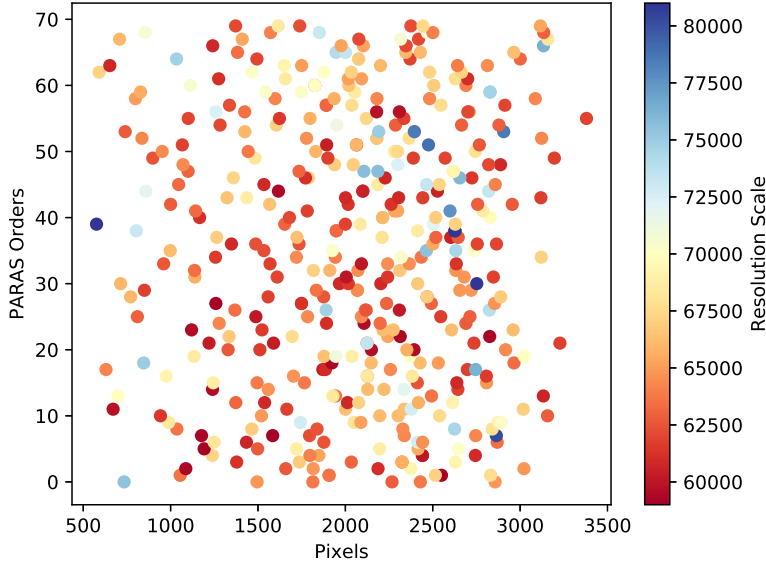}
\caption{The resolution of acquired UAr spectrum in various orders. It can be seen that the spectral resolution is in the range of 60000-80000. The color coding here represents the resolution and, the median resolution is 67000.} 
\label{fig:resolution}
\end{figure}

\begin{figure}[!ht]
\centering
\includegraphics[width=0.7\textwidth, trim= {0.0cm 0.0cm 0.0cm 0.0cm}]{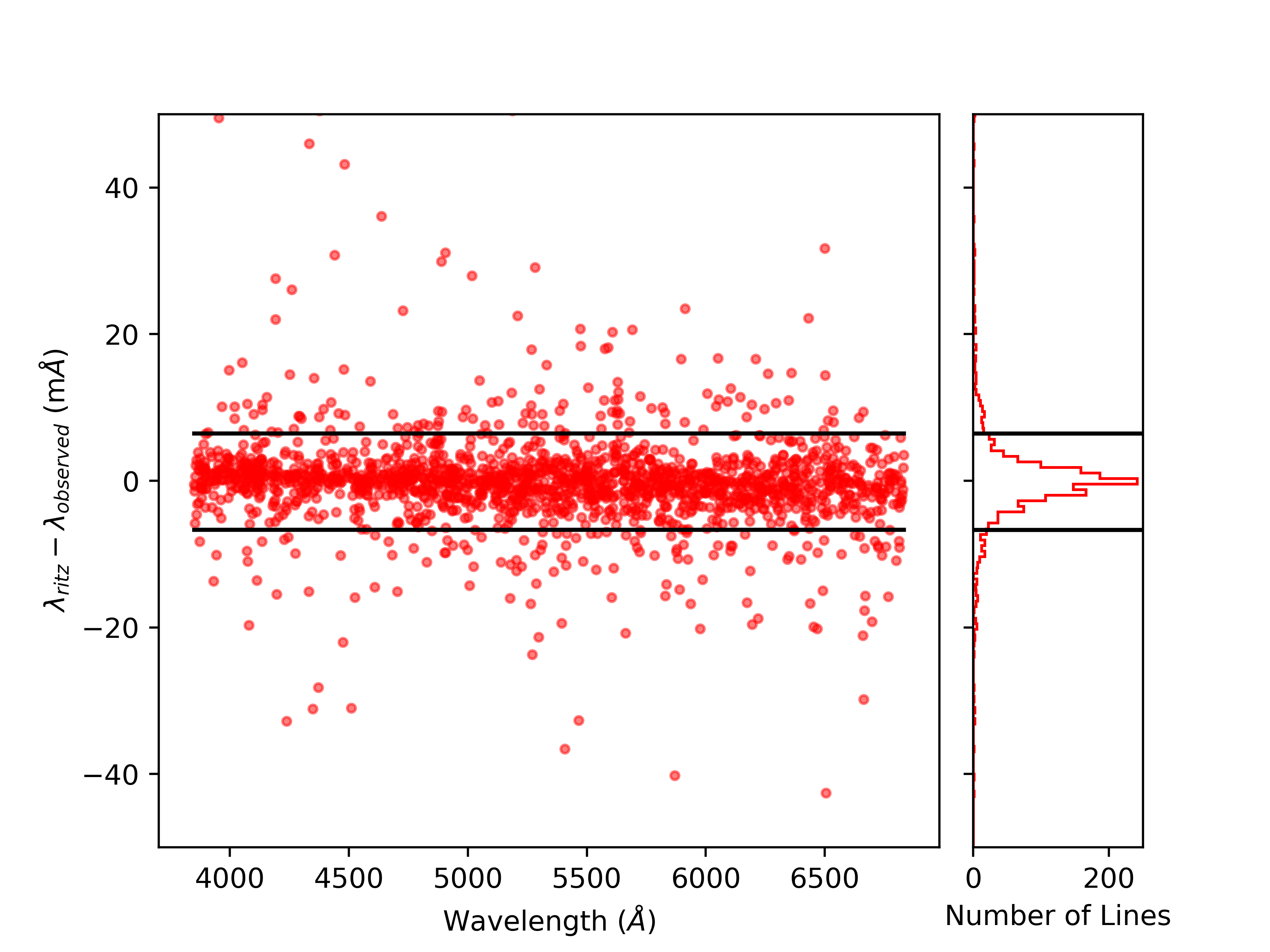}
\caption{{The distribution of differences between observed central wavelengths in this work and the Ritz wavelength. The left panel is the scatter plot of the distribution. The median and standard deviation ($1\sigma$) of the distribution are -0.15 and 2.1 m\AA, respectively. The right panel is the histogram of the same data. The two black lines represent the $\pm3\sigma$ limits of the distribution.}} \label{fig:diff_theo}
\end{figure}

In this way, the total number of lines selected {after passing through each of the above selection steps} was {1819}. We call this line list as the preliminary U line list. The instrumental resolution calculated using these lines by measuring their line widths is plotted in fig-\ref{fig:resolution}. The median resolution of the spectrograph is 67000.
{\section{The Final U line list}}\label{sec:ritz}
The preliminary line list possibly contains the lines which are not exactly the U transitions or for which energy levels are unknown.{ It is also possible that the UAr spectra may contain more Ar lines than the ThAr spectra, which are mentioned in \citenum{lovispepe2007} (see sec~\ref{sec:id_lines}). As stated before, we do not want to include any Ar line in the final line list; therefore, we compared all the lines from the preliminary line list with the possible theoretical transitions.} We calculated the Ritz wavelength using the energy levels for the first and the second transition of U found at actinides database \footnote{\url{http://www.lac.u-psud.fr/old-lac/lac/Database/Tab-energy/Uranium/}}, considering the selection rule for possible allowed electronic transitions\cite{thritz2014}. With the known energy levels, there are 96936 possible transitions in the wavelength range of 3809-6833 \AA. We calculated the difference between all our lines and the closest line from the theoretical line list, {and plotted its distribution in Fig~\ref{fig:diff_theo}. The median of the distribution is found to be -0.15 m\AA, with a standard deviation ($\sigma$ ) of 2.1 m\AA. {The standard error of median is 0.061 m\AA, which shows that the offset is considerable.} The observed lines that lie within the $3\sigma$ of the median of this distribution from their closest Ritz wavelength are selected for the final line list.} {The final line list contains a total of 1540 U lines.} The remaining {279} lines from preliminary line list are either U lines for unidentified energy levels or atomic lines for other elements that may be present as contamination in U. The identification of these lines is not in the scope of this paper, and we simply rejected these lines.\\

\begin{figure}[!ht]
\centering
\includegraphics[width=0.7\textwidth, trim= {0.0cm 0.0cm 0.0cm 0.0cm}]{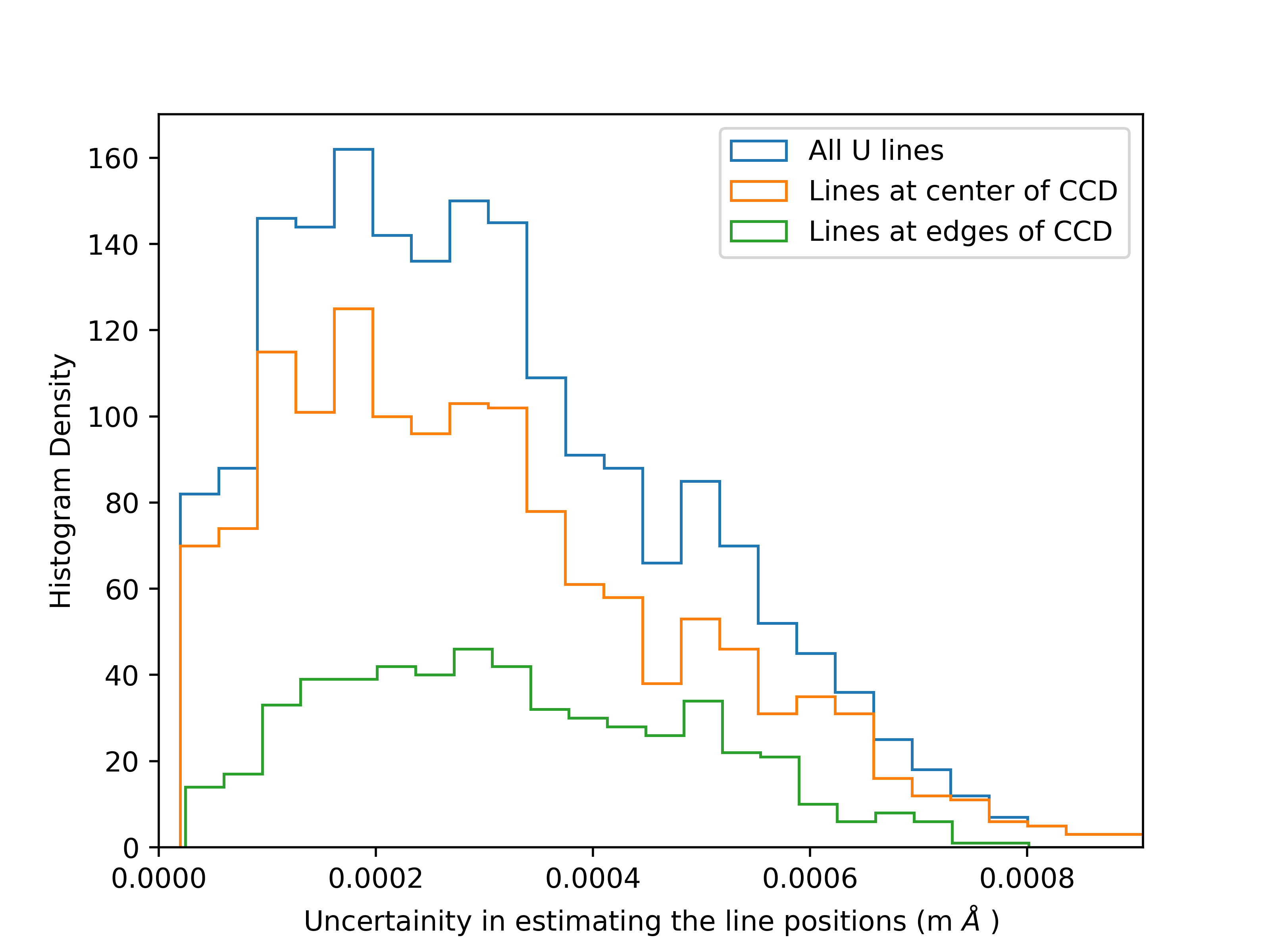}
\caption{The histogram plot for the uncertainties with which we find the positions of each line in the final line list in {m\AA, is plotted in blue. The mean error is 0.28 m\AA. The orange and green colored histograms show the error distributions of the lines falling at the central part of the CCD and the edges of the CCD with a mean of 0.27 m\AA, and 0.31 m\AA, respectively. } } \label{fig:err_uar}
\end{figure}

The finally selected lines are listed in Table-\ref{tab:line list}, which has 7 columns. 
{The uncertainties reported in the table are the errors associated with each line while determining the line position, computed from the covariance matrix using the {\tt MPFITPEAK} \cite{mpfit2009} Gaussian function written in {\tt IDL}. The major contributor in these errors is the photon shot noise corresponding to the line intensities and they range between 0.02 - 0.9 m\AA.} The final line list has an average measurement uncertainty of 15 m s$^{-1}$ (0.013 pixels or 0.28 m\AA). The strongest line has a measurement uncertainty of $\sim$1.2 m s$^{-1}$ (0.001 pixels or 0.02 m\AA), while the faintest line has been measured with a precision of $\sim$50 m s$^{-1}$ (0.043 pixels or 0.9 m\AA). It is also observed that the median photon noise errors of the lines falling at the edges are 14 \% higher than the lines falling at the center of the CCD. This is because that lines falling on the edges of the detector, where orders are stitched together may have slight asymmetric point spread function (PSF), which leads to higher errors. The various distributions of uncertainties are plotted in fig~\ref{fig:err_uar}. It is to note here that the Th lines used for initial wavelength calibration of the U lines has a median offset of $\sim$-1.5 m\AA\ from the absolute wavelength scale of the \citenum{lovispepe2007} line list, and that offset has been corrected before comparing the U lines (from preliminary U line list) with Ritz wavelengths.

\begin{figure}[!ht]
\centering
\includegraphics[width=0.7\textwidth, trim= {0.0cm 0.0cm 0.0cm 0.0cm}]{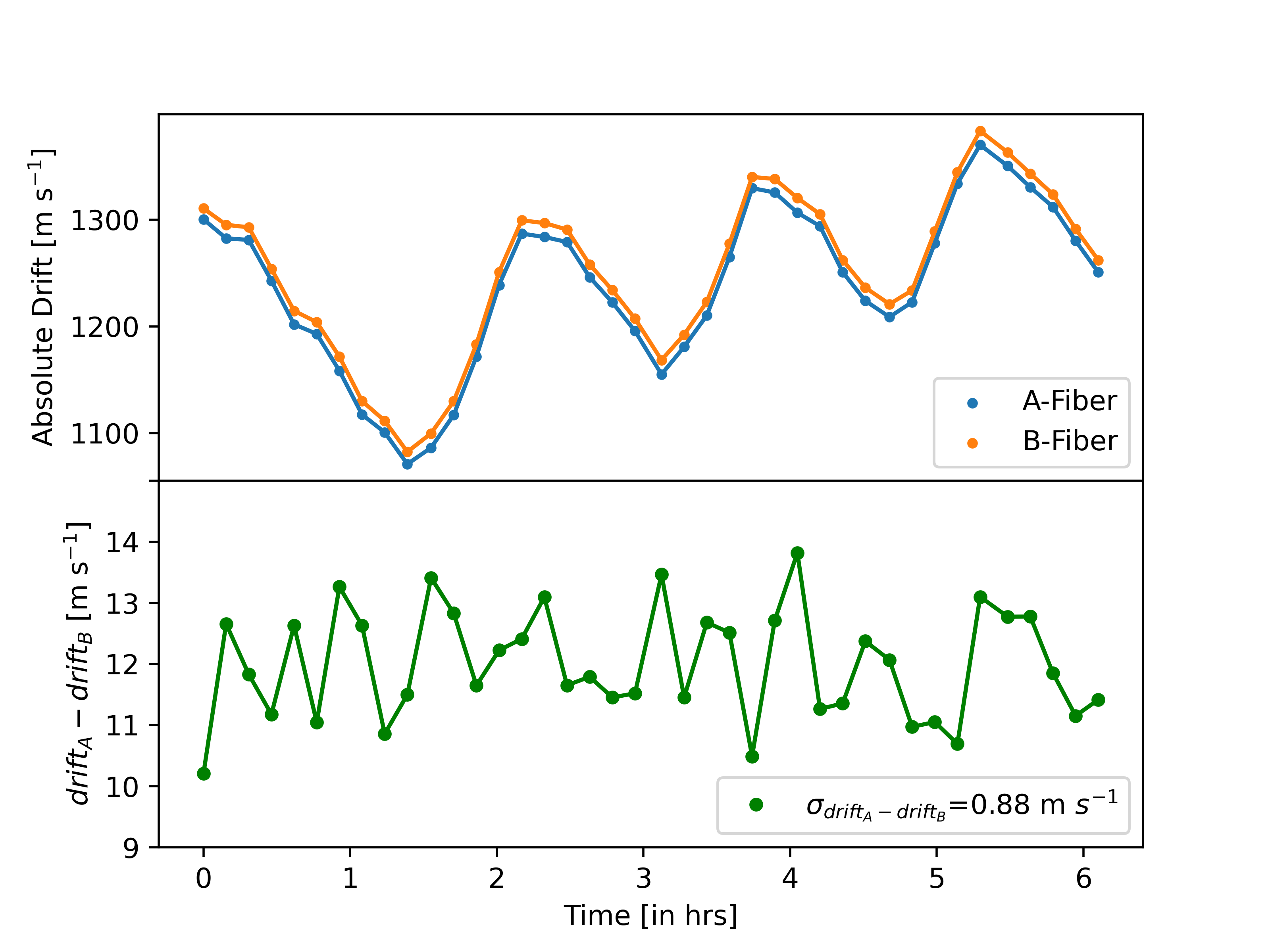}
\caption{{Upper Panel:- The figure shows the absolute drift in A and B fiber with respect to time. The dispersion in absolute drift is 75 m s$^{-1}$ in both fibers. It can be seen that both the fibers track each other very well.
Lower Panel:-The figure shows the inter-fiber drift between the A and B fiber for the course of 6.5 hrs in m s$^{-1}$, measured using UAr HCL's spectra acquired in both the fibers, wavelength calibrated with our new line list, listed in Table~\ref{tab:line list} . The average differential drift between the two fibers is $\sim$12 m s$^{-1}$ with a dispersion of 0.88 m s$^{-1}$ for the mentioned duration.} \label{fig:uar_uar}}
\end{figure}

\begin{figure}[!ht]
\centering
\includegraphics[width=0.7\textwidth, trim= {0.0cm 0.0cm 0.0cm 0.0cm}]{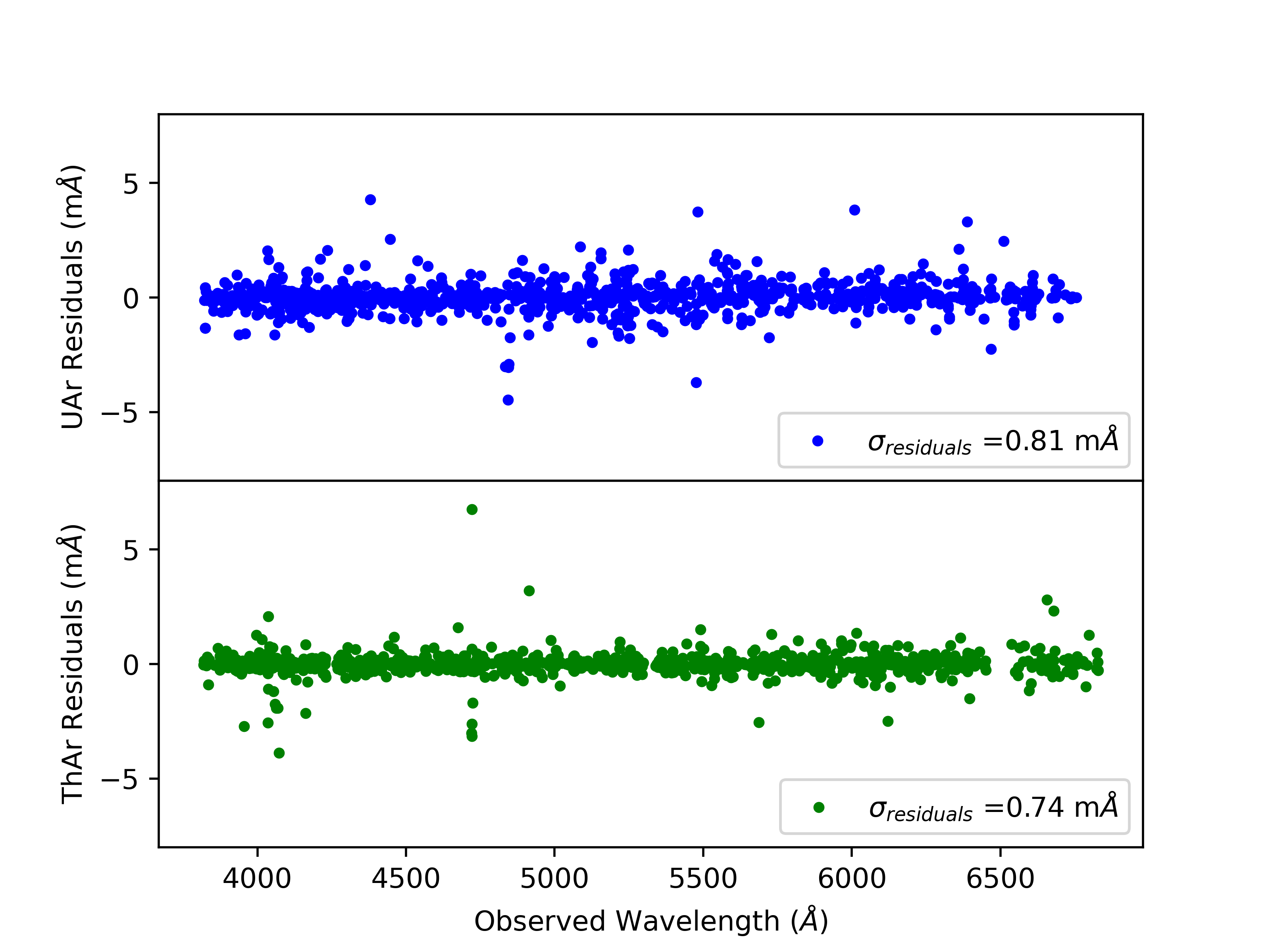}
\caption{{The plot here shows the residuals between the observed central wavelength of a random UAr spectrum and the central wavelength from the line list for UAr in the upper panel and the same for ThAr in the lower panel using Th lines. The dispersion of the residuals to the wavelength solution is 0.81 m\AA\space for UAr, while 0.74 m\AA \space for ThAr. The dispersion for UAr residuals is within the 3$\sigma$ of the average measurement uncertainty (Fig~\ref{fig:err_uar}). }} \label{fig:thar_uar}
\end{figure}

\begin{figure}[!ht]
\centering
\includegraphics[width=0.7\textwidth, trim= {0.0cm 0.0cm 0.0cm 0.0cm}]{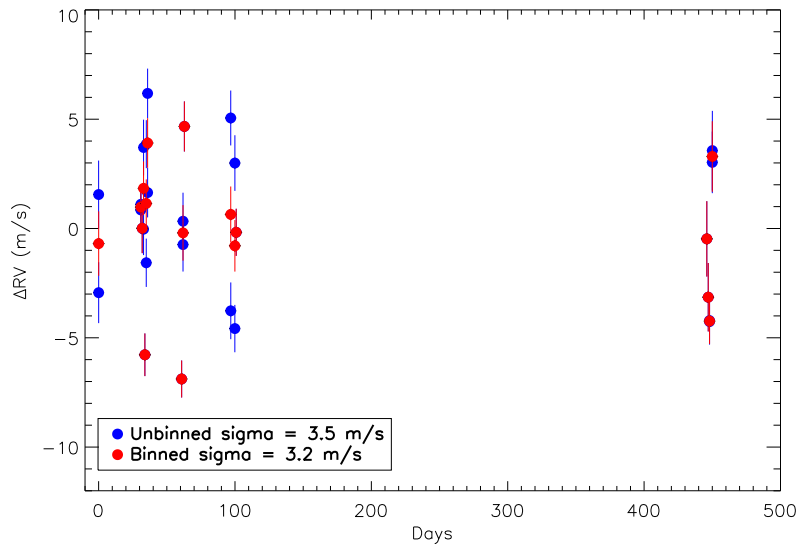}
\caption{The plot showing relative RVs of standard star HD55575 for the span of 450 days. The blue colored data points are the RV from each observed spectra calibrated with the UAr spectra, while the red ones are the binned RV datapoints, average to each night. The dispersion of binned and unbinned RVs is 3.5 and 3.2 \si{\meter\per\second}, respectively.} \label{fig:hd55575}
\end{figure}

{We have also compared our final line list with the P80 line list. The P80 line list is the only previously published line list which covers the whole wavelength range used in our work. They used Fourier Transform Spectrograph (FTS) for the line identification work, and reported around 3700 lines in the PARAS wavelength range. We found that $\sim$90\% of the lines in the final line list are from P80, and the remaining are newly identified in this work. These newly identified lines are flagged as 'RS21' in the line list, while the remaining have been flagged as 'P80' (see Table-\ref{tab:line list}). In the 1980s, Photo Multiplier Tubes (PMTs) were generally used as detectors, while we use a CCD as a detector, which has much better quantum efficiency than the former. It results in better sensitivity and thus, we could detect the fainter lines missed in P80 work. Also, new energy levels have been identified after the P80 work, found at the actinides energy database, are included in this work  
{\footnote{\url{http://www.lac.u-psud.fr/old-lac/lac/Database/Tab-energy/Uranium/U1-ref.html}}}{\footnote{\url{http://www.lac.u-psud.fr/old-lac/lac/Database/Tab-energy/Uranium/U2-ref.html}}}. The number of lines in our line list is less than the P80 line list as our resolution is much lesser than the FTS $(R=600,000)$, and our line list only contains the lines, which are clearly resolved and their line positions are precisely determined. Besides that, we have found that there is no significant offset in the absolute wavelength scale in our line list from P80 line list. The main advantage of our line list is that the photon noise uncertainties in determining the line positions are much lesser than the P80. The P80 line list determined the line positions with an {average} precision of 2 m\AA, while the median error of our line list is 0.28 m\AA, which is $\sim$ 7 times better than the P80 line list errors.}
 {\section{Results and Discussions}\label{sec:res_des}
\subsection{Performance of the instrument using UAr HCL}\label{sec:res_uar}}
{We measured and found that the fundamental precision achievable with the UAr spectra is 20 cm s$^{-1}$, while it is 18 cm s$^{-1}$ with the ThAr spectra in PARAS, calculated using the algorithm explained in \citenum{Bouchy2001}.}
To check the stability of the UAr HCL with the instrument and to track the instrumental drift, {a series of continuous UAr-UAr exposures is acquired for $\sim$6.5 hours as described in section~\ref{sec:observations}}. {A template of the U lines from the final line list is used for the wavelength calibration of these spectra. The template is created in the same manner as described in section~\ref{sec:data_reduction}. }
This {template} is replaced with the previously used Th lines {template} in the PARAS data analysis {framework}, and the wavelength calibration of the spectra is done using the same methodology as described in section~\ref{sec:data_reduction}. For an order, the new pixel position of each line belonging to that particular order is estimated and, then a {third-order polynomial} is fitted over these new positions and the central wavelength of the lines in the {template}. Then that polynomial solution is applied over the whole order. The same procedure has been iterated for all the 70 orders. In this way, the wavelength calibration of all the UAr-UAr spectra was done. 
The instrumental drift is calculated by cross-correlating each of these UAr spectra with a binary mask of U lines used in the wavelength calibration. The binary mask is defined as a template spectrum of U, which consists of several box-shaped or rectangular functions, each of them centered at U line position and has a constant value over the width associated with that line and is zero elsewhere.
We have plotted the instrumental  in fig~\ref{fig:uar_uar}. The upper panel shows the absolute drift in both the fibers, while the lower panel depicts the relative drift between the two fibers with respect to time. {Since both the fibers (fiber A \& fiber B) pass through the same opto-mechanical elements and see same minute changes in pressure and temperature, therefore the drift in each fiber should be identical \cite{paras2014}. We find that both the fibers drift identically (see fig~\ref{fig:uar_uar}, upper panel) and the 1$\sigma$ scatter in the absolute drift is $\sim$75 m s$^{-1}$ in fiber A as well as in fiber B. However, there is a better way to represent the achievable RV precision of the spectrograph using the differential drift (or inter-fiber drift) between both the fibers\cite{paras2014}. In PARAS, we use the simultaneously acquired calibration lamp's spectrum for the wavelength calibration of a stellar spectrum in order to correct the absolute instrumental drifts. Also, the inter-fiber drift is corrected using the UAr-UAr exposures acquired just before and after the stellar spectrum and that minimizes the wavelength calibration errors. Therefore, the 1$\sigma$ scatter of this inter-fiber drift exhibits the RV precision due to the wavelength calibration errors. Here, the 1$\sigma$ of the differential drift ($\sigma_{driftA-driftB}$) is found to be 0.88 m s$^{-1}$. Earlier, with the ThAr HCL, this 1$\sigma$ scatter was 0.92 m s$^{-1}$\cite{paras2014}.} These results are comparable as the 1$\sigma$ scatter of the differential drift from both the HCLs is in good agreement with each other. It is to note here that the typical dispersion in the differential drift with a UAr HCL on a night is $\leq$ 1 m s$^{-1}$, and it also represents the wavelength calibration precision. This demonstrates that the UAr HCLs can replace the ThAr HCLs in PARAS-like spectrographs for achieving the wavelength calibration precision up to 1 m s$^{-1}$. {PARAS-like spectrographs here refer to the spectrographs with similar spectral resolution (R $\leq$ 70,000) as PARAS, similar pressure and temperature stability as described in \citenum{paras2014} and  uses the similar calibration system (like ThAr or UAr HCL). } \\
Furthermore, we checked the residuals around the wavelength solution of a UAr spectrum from the line positions listed in the U lines {template}. The same has been done for the ThAr spectra using the Th lines {template}. We have plotted the residuals in fig~\ref{fig:thar_uar}. The standard deviation of the residuals for ThAr and UAr spectrum is found to be 0.74 and 0.81 m\AA, respectively and, the difference between the two is within the photon noise error limit. {The dispersion of the residuals from UAr spectrum is within the 3-$\sigma$ of the average measurement uncertainty associated with the final U line list.}
{\subsection{The RV precision achievable on HD55575 with UAr HCL}}\label{sec:res_star}
After successfully testing the HCL with the instrument and off-sky drift measurements, we did an on-sky test with a well-known RV standard star, HD55575\cite{hd55575}, which has shown the RV dispersion of 3.4 \si{\meter\per\second} over the span of 118 days with SOPHIE+ {\cite{hd55575}} and 3.1 \si{\meter\per\second} with PARAS using Th lines\cite{epic201}, over the span of 420 days. We observed HD55575 for the span of $\sim$ 450 days, acquiring 27 spectra, simultaneously with the UAr HCL. The same U-lines {template} is used for the wavelength calibration of the extracted stellar spectra, which is used in Sec~\ref{sec:res_uar}. We imposed the wavelength solution of the simultaneously acquired UAr spectra over the stellar spectra.
The instrumental drift in both the fibers is corrected using the UAr-UAr exposures taken just before and after the stellar spectra. {We computed the RVs by cross-corelating the stellar spectra against the template stellar mask of a G2-type star.} The mean absolute stellar RV using UAr HCL is found to be $85.7597\pm0.001$ Km s$^{-1}$, while it was $85.7498\pm0.001$ Km s$^{-1}$ for the stellar spectra acquired simultaneously with ThAr HCL. This corresponds to an absolute RV offset of $\sim$10 m $s^{-1}$ {(0.18 m\AA\space at 5500 \AA)} between the two datasets, which can be treated as an instrumental offset between the two HCLs. {This offset is very close to the offset found between our final line list and the corresponding Ritz wavelengths for U lines (Sec~\ref{sec:ritz}) and it is a supporting evidence of consistency between wavelength calibration systemics using Th and U lines.} The relative RVs for HD55575 are plotted in fig~\ref{fig:hd55575}, and the RV dispersion for the star is found to be 3.2 \si{\meter\per\second} by nightly binning the data for the whole span of 450 days. {Previously, with the use of ThAr HCL, we have achieved a RV dispersion of 3.1 \si{\meter\per\second} over 391 days\cite{epic201}. It shows that the long-term RV dispersion measured here with UAr HCL is in good agreement with the previous measurements using ThAr HCL.}\\
We have shown that the same level of RV precision on stars as well as the same level of wavelength calibration precision is achievable using the UAr HCLs compared to the ThAr HCLs in the PARAS spectrograph. Thus, ThAr HCLs can be replaced with the UAr HCLs for the precise wavelength calibration of a high-resolution spectrum in the visible region, in PARAS-like spectrographs for exoplanet detection work and to achieve the RV precision of 1-3 m s$^{-1}$.
{\section{Future works}}
{We are planning to use the UAr HCL in our upcoming high-resolution spectrograph, PARAS-2\cite{PARAS-2} (R $\sim$ 100,000) for instrumental characterization and wavelength calibration. The spectrograph will be attached to the upcoming PRL 2.5 meter telescope at Mount Abu Observatory, India for a goal of achieving sub-1 m s$^{-1}$ level of RV precision. }\\ 

\section{Summary}
{We have shown here the precise wavelength calibration of a high-resolution spectrum using U lines, acquired with the PARAS spectrograph, in the wavelength range of 3809-6833 \AA. We initially identified a total of 4500 line features in the UAr HCL's spectra and then filtered them through various selection steps to keep only well-resolved, well-separated, and unblended 1819 U lines. Then, we compared these lines with the closest Ritz wavelength and considered only 1540 lines for the final line list, discarding the lines with a significantly higher difference from Ritz wavelength. The comparison showed an offset of {-0.15 m\AA}, with a standard deviation of 2.1 m\AA\space and the average wavelength measurement uncertainty of our final line list is 0.28 m\AA. We identified 160 new U transitions in this work other than the previous work of P80, and the average uncertainty in estimating the line positions in our work was found much better than that of P80. Finally, the line list was included in the PARAS data analysis framework to check the instrumental stability and RV precision. The typical dispersion of residuals in wavelength solution of a UAr spectrum is measured to be 0.8 m\AA, which is within 3-$\sigma$ of mean errors in estimating the line positions. The 1$\sigma$ scatter of the inter-fiber drift is found to be 88cm s$^{-1}$ over 6.5 hours. {The stellar spectra wavelength calibrated using U lines and Th lines shows an absolute RV offset of $\sim$10 m $s^{-1}$ (0.18 m\AA\space at 5500 \AA). This offset is a supporting evidence of consistency between wavelength calibration systemics using Th and U lines as it is in agreement with the offset our linelist from absolute wavelength scale.} We measured the RV dispersion ($\sigma_{RV}$) of 3.2 m s$^{-1}$ on a RV reference star HD55575 over the course of 450 days, using of U lines. These results are in good agreement with the previous results derived using wavelength calibration of ThAr HCL. Thus, we have demonstrated that the on-sky and off-sky acquired spectra, wavelength calibrated with the simultaneously acquired UAr HCL spectra, shows good stability in the RV and, therefore, can be used as a replacement for the ThAr HCL in the high-resolution astronomical spectrographs for precision RV measurements at a level of 1-3 m s$^{-1}$.}
\\

\section{Acknowledgments}
The PARAS spectrograph is fully funded and being supported by Physical Research Laboratory (PRL), which is part of Department of Space, Government of India. RS and AC would like to thank Director, PRL for his support and encouragement. RS is thankful to Priyanka Chaturvedi from TLS, Tautenburg, Germany for her valuable suggestions in improving the script and efforts in developing the various codes for analysis of the PARAS spectra. RS also acknowledges the help from Vishal Shah, Kapil Kumar, Neelam JSSV Prasad, Kevi Kumar, Ashirbad Nayak, Dishendra, Akanksha Khandelwal and Mount Abu Observatory staff at the time of observations. AC is grateful to Suvrath Mahadevan and Arpita Roy from Pennsylvania University, USA, for their tremendous efforts in developing the PARAS data pipeline in 2014.

\bibliography{astro_citations}   

\begin{thebibliography}{10}

\bibitem{Rouan1998}
D.~{Rouan}, A.~{Baglin}, E.~{Copet}, {\em et~al.}, ``{The Exosolar Planets
  Program of the COROT satellite},'' {\em Earth Moon and Planets} {\bf 81},
  79--82  (1998).

\bibitem{Borucki2010}
W.~J. {Borucki}, D.~{Koch}, G.~{Basri}, {\em et~al.}, ``{Kepler
  Planet-Detection Mission: Introduction and First Results},'' {\em Science}
  {\bf 327}, 977  (2010).

\bibitem{Ricker2015}
G.~R. {Ricker}, J.~N. {Winn}, R.~{Vanderspek}, {\em et~al.}, ``{Transiting
  Exoplanet Survey Satellite (TESS)},'' {\em Journal of Astronomical
  Telescopes, Instruments, and Systems} {\bf 1}, 014003  (2015).

\bibitem{Swasp}
D.~L. Pollacco, I.~Skillen, A.~C. Cameron, {\em et~al.}, ``The {WASP} project
  and the {SuperWASP} cameras,'' {\em Publications of the Astronomical Society
  of the Pacific} {\bf 118}, 1407--1418  (2006).

\bibitem{KELT}
J.~Pepper, R.~W. Pogge, D.~L. DePoy, {\em et~al.}, ``The kilodegree extremely
  little telescope ({KELT}): A small robotic telescope for large-area synoptic
  surveys,'' {\em Publications of the Astronomical Society of the Pacific} {\bf
  119}, 923--935  (2007).

\bibitem{EXPRES2016}
C.~{Jurgenson}, D.~{Fischer}, T.~{McCracken}, {\em et~al.}, ``{EXPRES: a next
  generation RV spectrograph in the search for earth-like worlds},'' in {\em
  Ground-based and Airborne Instrumentation for Astronomy VI},  C.~J. {Evans},
  L.~{Simard}, and H.~{Takami}, Eds., {\em Society of Photo-Optical
  Instrumentation Engineers (SPIE) Conference Series} {\bf 9908}, 99086T
  (2016).

\bibitem{espresso2014}
F.~{Pepe}, P.~{Molaro}, S.~{Cristiani}, {\em et~al.}, ``{ESPRESSO: The next
  European exoplanet hunter},'' {\em arXiv e-prints,} , arXiv:1401.5918
  (2014).

\bibitem{wilken2012}
T.~{Wilken}, G.~L. {Curto}, R.~A. {Probst}, {\em et~al.}, ``{A spectrograph for
  exoplanet observations calibrated at the centimetre-per-second level},'' {\em
  {Nature,}} {\bf 485}, 611--614  (2012).

\bibitem{hpf2019}
A.~J. {Metcalf}, T.~{Anderson}, C.~F. {Bender}, {\em et~al.}, ``{Stellar
  spectroscopy in the near-infrared with a laser frequency comb},'' {\em
  {Optica,}} {\bf 6}, 233  (2019).

\bibitem{Curto2012}
G.~{Lo Curto}, L.~{Pasquini}, A.~{Manescau}, {\em et~al.}, ``{Astronomical
  Spectrograph Calibration at the Exo-Earth Detection Limit},'' {\em The
  Messenger} {\bf 149}, 2--6  (2012).

\bibitem{Kerber2007}
F.~{Kerber}, G.~{Nave}, C.~J. {Sansonetti}, {\em et~al.}, ``{The Spectrum of
  Th-Ar Hollow Cathode Lamps in the 900-4500 nm Region: Establishing Wavelength
  Standards for the Calibration of VLT Spectrographs},'' in {\em The Future of
  Photometric, Spectrophotometric and Polarimetric Standardization},
  C.~{Sterken}, Ed., {\em Astronomical Society of the Pacific Conference
  Series} {\bf 364}, 461  (2007).

\bibitem{paras2014}
A.~{Chakraborty}, S.~{Mahadevan}, A.~{Roy}, {\em et~al.}, ``{The PRL Stabilized
  High-Resolution Echelle Fiber-fed Spectrograph: Instrument Description and
  First Radial Velocity Results},'' {\em {PASP,}} {\bf 126}, 133  (2014).

\bibitem{epic201}
A.~{Chakraborty}, A.~{Roy}, R.~{Sharma}, {\em et~al.}, ``{Evidence of a
  Sub-Saturn around EPIC 211945201},'' {\em {The Astronomical Journal,}} {\bf
  156}, 3  (2018).

\bibitem{HARPS_longterm}
X.~{Dumusque}, F.~{Pepe}, C.~{Lovis}, {\em et~al.}, ``{An Earth-mass planet
  orbiting {\ensuremath{\alpha}} Centauri B},'' {\em {Nature,}} {\bf 491},
  207--211  (2012).

\bibitem{Nave2018}
G.~{Nave}, F.~{Kerber}, E.~A. {Den Hartog}, {\em et~al.}, ``{The dirt in
  astronomy's genie lamp: ThO contamination of Th-Ar calibration lamps},'' {\em
  {Proceedings of the SPIE,}} {\bf 10704}, 1070407  (2018).

\bibitem{Redman2012}
S.~L. {Redman}, G.~G. {Ycas}, R.~{Terrien}, {\em et~al.}, ``{A High-resolution
  Atlas of Uranium-Neon in the H Band},'' {\em {The Astrophysical Journal
  Supplement Series,}} {\bf 199}, 2  (2012).

\bibitem{carmenes18}
L.~F. {Sarmiento}, A.~{Reiners}, P.~{Huke}, {\em et~al.}, ``{Comparing the
  emission spectra of U and Th hollow cathode lamps and a new U line list},''
  {\em {Astronomy and Astrophysics,}} {\bf 618}, A118  (2018).

\bibitem{palmer80}
B.~A. {Palmer}, R.~A. {Keller}, and J.~{Engleman}, Rolf, ``{Atlas of uranium
  emission intensities in a hollow cathode discharge},'' {\em Los Alamos
  Scientific Laboratory,} {\bf 12}, Report number LA--8251--MS  (1980).

\bibitem{reduce}
N.~E. {Piskunov} and J.~A. {Valenti}, ``{New algorithms for reducing
  cross-dispersed echelle spectra},'' {\em {Astronomy and Astrophysics,}} {\bf
  385}, 1095--1106  (2002).

\bibitem{mpfit2009}
C.~B. {Markwardt}, {\em {Non-linear Least-squares Fitting in IDL with MPFIT}},
  vol.~411 of {\em {Astronomical Society of the Pacific Conference Series,}},
  251  (2009).

\bibitem{lovispepe2007}
C.~{Lovis} and F.~{Pepe}, ``{A new list of thorium and argon spectral lines in
  the visible},'' {\em {Astronomy and Astrophysics,}} {\bf 468}, 1115--1121
  (2007).

\bibitem{thritz2014}
S.~L. {Redman}, G.~{Nave}, and C.~J. {Sansonetti}, ``{The Spectrum of Thorium
  from 250 nm to 5500 nm: Ritz Wavelengths and Optimized Energy Levels},'' {\em
  {The Astrophysical Journal Supplement Series,}} {\bf 211}, 4  (2014).

\bibitem{Bouchy2001}
F.~{Bouchy}, F.~{Pepe}, and D.~{Queloz}, ``{Fundamental photon noise limit to
  radial velocity measurements},'' {\em {Astronomy and Astrophysics,}} {\bf
  374}, 733--739  (2001).

\bibitem{hd55575}
F.~{Bouchy}, R.~F. {D{\'\i}az}, G.~{H{\'e}brard}, {\em et~al.}, ``{SOPHIE+:
  First results of an octagonal-section fiber for high-precision radial
  velocity measurements},'' {\em {Astronomy and Astrophysics,}} {\bf 549}, A49
  (2013).

\bibitem{PARAS-2}
A.~Chakraborty, N.~Thapa, K.~Kumar, {\em et~al.}, ``{PARAS-2 precision radial
  velocimeter: optical and mechanical design of a fiber-fed high resolution
  spectrograph under vacuum and temperature control},'' in {\em Ground-based
  and Airborne Instrumentation for Astronomy VII},  C.~J. Evans, L.~Simard, and
  H.~Takami, Eds.,  {\bf 10702}, 1967 -- 1978, International Society for Optics
  and Photonics, SPIE  (2018).

\end{thebibliography}
\bibliographystyle{spiejour}

\begin{landscape}
\begin{longtable}{c c c c c c c} 
\caption{The final line list of uranium from PARAS spectrograph.}
\label{tab:line list}\\\hline\hline    
Observed Wavelength & Ritz Wavelength & Relative Flux & Uncertainty & Offset from Ritz wavelength & Species & Source\\
\hline
\AA & \AA & ADU & m\AA & m\AA & - & -\\
\hline\hline
\endfirsthead
\hline
Observed Wavelength & Ritz Wavelength & Relative Flux & Uncertainty & Offset from Ritz wavelength & Species & Source\\
\hline
\AA & \AA & ADU & m\AA & m\AA & - & -\\
\hline\hline
\endhead 
\hline
\multicolumn{3}{r@{}}{continued \ldots}\\
\endfoot
\hline
\endlastfoot
 3809.4663 & 3809.4617 &   	    6153 & 	  0.2709 &    4.6000 &  UII  &  RS21 \\
 3821.9472 & 3821.9467 &        2900 &    0.3183 &    0.5353 &  UI   &  RS21   \\
 3824.5337 & 3824.5374 &        4240 &    0.2671 &   -3.7387 &  UI   &  RS21   \\
 3829.7954 & 3829.7935 &        2885 &    0.3030 &    1.9979 &  UI   &  RS21   \\
 3831.8545 & 3831.8509 &        2295 &    0.3486 &    3.5569 &  UI   &  RS21   \\
 3835.2266 & 3835.2240 &        4017 &    0.2662 &    2.6342 &  UI   &  RS21   \\
 3839.6262 & 3839.6267 &       16468 &    0.1219 &   -0.5369 &  UI   &  RS21   \\
 3844.9362 & 3844.9361 &        4423 &    0.2621 &    0.0783 &  UI   &  RS21   \\
\end{longtable}
\end{landscape}
\end{document}